\documentclass[twocolumn,aps,pre,superscriptaddress,showpacs]{revtex4}
\usepackage{amsmath,bm}
\usepackage{graphicx}
\usepackage{dcolumn}
\usepackage{multirow}
\begin{document}
\draft
\title{Network Topology of the Austrian Airline Flights }

\author{Ding-Ding Han \email{Corresponding author. E-Mail: ddhan@ee.ecnu.edu.cn}}
\author{ Jiang-Hai Qian }
\author{ Jin-Gao Liu }



\affiliation{\footnotesize College of Information, East China
Normal University, Shanghai 200062, China}


\date{\today}
\nopagebreak

\begin{abstract}

The information of the Austrian airline flights was collected and
quantitatively analyzed by the concepts of complex network. It
displays some features of small-world networks, namely large
clustering coefficient and small average shortest-path length. The
degree distributions of the networks reveal power law behavior
with exponent value of 2 $\sim$ 3 for the small degree branch but
a flat tail for the large degree branch. Similarly, the flight
weight distributions show power-law behavior for the small weight
branch. Furthermore, we found that the clustering coefficient $C$,
0.206, of this flight network is greatly larger than that of a
random network, 0.01, which has the same numbers of the airports
($N$) and mean degree ($\langle k \rangle$), and the diameter $D$,
2.383, of the flight network is significantly smaller than the
value of the same random network, 18.67. In addition, the
degree-degree correlation analysis shows the network has
disassortative behavior, i.e.  the large airports are likely to
link to smaller airports. Furthermore, the clustering coefficient
analysis  indicates that the large airports reveal the
hierarchical organization.



\end{abstract}

\maketitle

Network behaviors emerge across many interdisciplinary sciences
and attract the interests from many researchers in different
research fields. Network is usually a set of items, which we will
call vertices or sometimes nodes, with connections between them,
called edges. Systems with the form of networks are distributed
over the world. Examples \cite{BA-RMP,Dorogovtsev,Newm}  include
the Internet, the World Wide Web (WWW), social networks of
friends, networks of business relations between companies, neutral
network, metabolic networks, food webs, distribution networks such
as blood vessels or postal delivery routes, networks of citations
between papers,  networks of paper collaborators, network of
publication download frequency \cite{Han}, and traffic
transportation networks \cite{cai,cai2,cai3,Railway,Zhou,Xu} and
many others \cite{He,Wang,Wu,Ouyang,Liu1,Chen}. Even in
microscopic scale, such as in nuclear fragmentation produced in
heavy ion collisions, the hierarchial power-law distribution of
nuclear fragments emerges by Ma's nuclear Zipf-type plots around
the nuclear liquid gas phase transition, which also shows a
similar character to the scale free network \cite{Ma,Ma2}.

In a pioneering work of Barabasi and Albert, they found that the
degree of node of Internet routes, URL (universal resource
locator) - linked networks in the WWW  satisfies the power-law
distribution \cite{BA,BA2}, also called as the scale-free
networks. The degree distribution of a scale-free network is a
power law,
\begin{equation}
P_k \sim k^{-\gamma},
\end{equation}
where $\gamma$ is called the degree exponent. In addition, there
are other two main topological structures of complex networks:
random-graph models \cite{rand} and  small-world networks
\cite{Wat}.  The research on random graphs was initiated by
Erd\"os and R\'enyi in the 1950s. It has shown that the degree
distribution of a random graph is of Poisson type, $P(k) =
e^{-\langle k \rangle}\langle k \rangle/k!$. Small-world networks
are somehow in between regular networks and random ones
\cite{Koc}.

In this work we study the flight network affiliated with Austrian
Airline company to shed light on understanding the real
topological structure and inherent laws of flight network design
by a specific airline company. Some features of the flight network
will be compared with those of the above three categories of
networks. To this end, we would like to check the similarities and
differences among possibly different networks. Some studies have
been performed for the flight networks, such as for international
transportation airport networks by the foreign colleagues
\cite{Pnas,Guimera,Colizza} as well as the US and China flight
networks by Cai's group \cite{cai,cai2}. Some interesting features
have been demonstrated for such flight networks, such as the
small-world property: high clustering coefficient, small diameter
and hierarchical structure. However, our present work is different
in motivation and results. The previous flight network involved in
a whole national or international  airport networks, which did not
care about the detailed information of the fights which were
operated by a specific airline company. These national- or
world-wide flight networks are large scale
\cite{Pnas,Guimera,Colizza} but they are the result from
collective role by the various Airline company networks.
Therefore, it is of interesting to survey  a particular airline
flight network instead of a whole national or international- wide
flight network. Based upon this motivation, we will investigate a
smaller network which was composed by the flights affiliated with
a specific airline in the present work. As an example, we have
investigated the flight network of a central European airline
company, Austrian Airline. The flight information is available in
the web page, http://www.aua.com/

In the flight network, the airports can be represented by the
vertices and  the flights connecting two airports by edges. In the
previous studies \cite{Pnas,Guimera,Colizza,cai,cai2}, some
features of the structure of flight networks have been recognized:
(1) the network is directional. All the flights are directed,
sorted as outgoing and incoming. (2) the network has weight. To
reflect how busy a certain line is, it is important to record the
exact number of flights between any given airport $i$ and $j$
\cite{cai}, even to record the seat numbers available in different
flights \cite{Pnas}.  (3) the network may be a little different
day by day in a whole week. Hence, the weekly flight information
partially involves the information on evolution of the flight
network. Our data contain a whole week information of around $N
\sim$ 134 airports and 9560 flights. The detailed numbers of the
airports and flights are listed in Table I. For the  number of
flights, it is the largest on Monday and the smallest on Saturday.

The paper is organized as follows. First we present a sample of
the flight network in Friday and its degree distribution. Then we
give the results of the flight weight distributions, of the
clustering coefficient, of the diameter and of the assortative
coefficient, respectively. Finally a summary is given.

The vertex degree  distribution function $P_k$ gives the
probability that a randomly selected vertex has exactly $k$ edges
\cite{Bol}. Figure \ref{fig_netw} shows a topological structure of
the Austrian airline flight network on Friday, where each airport
is expressed by a node and the flights are connected by the lines
between two nodes. The Vienna airport is the dominative airport
operated by the Austrian airlines, which has naturally the largest
amount of edges. There are several major airports, such as Paris,
Frankfurt etc, which have frequent flights to connect with other
small airports operated by the Austrian Airlines. For comparison,
the Erd\"os and R\'enyi-type random network structure which has
the same vertices $N$ = 136 and mean degree $\langle k \rangle$ =
1.31 is also plotted in the figure. The obvious different
topological structure is there.

Three kinds of degree distributions, namely  $P_k(in)$, $P_k(out)$
and $P_k(all)$ are shown in Figure \ref{fig_degree}. $P_k(in)$ and
$P_k(out)$ represent the frequencies of incoming and outgoing of
flights, respectively. $P_k(all)$ is used when we do not
distinguish outgoing and incoming flights, i.e. it is just the
degree number which is regardless wether the flight is outgoing,
incoming, or both of them. Note that the present degree
distribution is not cumulative distribution as done in
Ref.~\cite{cai,cai2}. Even though the statistical fluctuation
could keep large in degree distribution in comparison with the
cumulative distribution, the distribution can give the direct
probability how many Austrian Airline flights are coming or taking
off. Two branches are seen in Figures \ref{fig_degree}(a) and
\ref{fig_degree}(b): the first one follows the power-law $P_k \sim
k^{-\gamma}$ when $k < 7$ and the second one is the flat tail
distribution when $k \geq 7$, which is basically related to some
largest airports which serve for Austrian Airlines. This behavior
can be partially attributed to different mechanisms between small
airports and large airports. For an example, they have different
growth rates since the construction of small airports or the
flight line extension to small airports by the Austrian Airlines
is much easier and faster than that of large airports. In the
following, we can extract the exponents of the degree distribution
for small airports which Austrian Airline flights cover. When $k <
7$, the mean weekly value of $\gamma_{in}$, $\gamma_{out}$ and
$\gamma_{all}$ correspond to 2.61, 2.63 and 2.47; Exponents in
each day in Figure \ref{fig_degree} are listed in Table I. The
average degree of the flight network is given by $\langle k
\rangle = \frac{1}{ N} \sum_i k_i$. The average $\langle k
\rangle_{all}$ = 1.30 . That means each airport is linked to 1.3
other airports for the flights affiliated with Austrian Airlines.
Similarly, $\langle k_{in} \rangle$ = 1.279 and $\langle k_{out}
\rangle$ = 1.277. In details, $\langle k \rangle$ on each day are
listed in Table I.

\begin{widetext}

\begin{figure}
\resizebox{20.2pc}{!}{\includegraphics{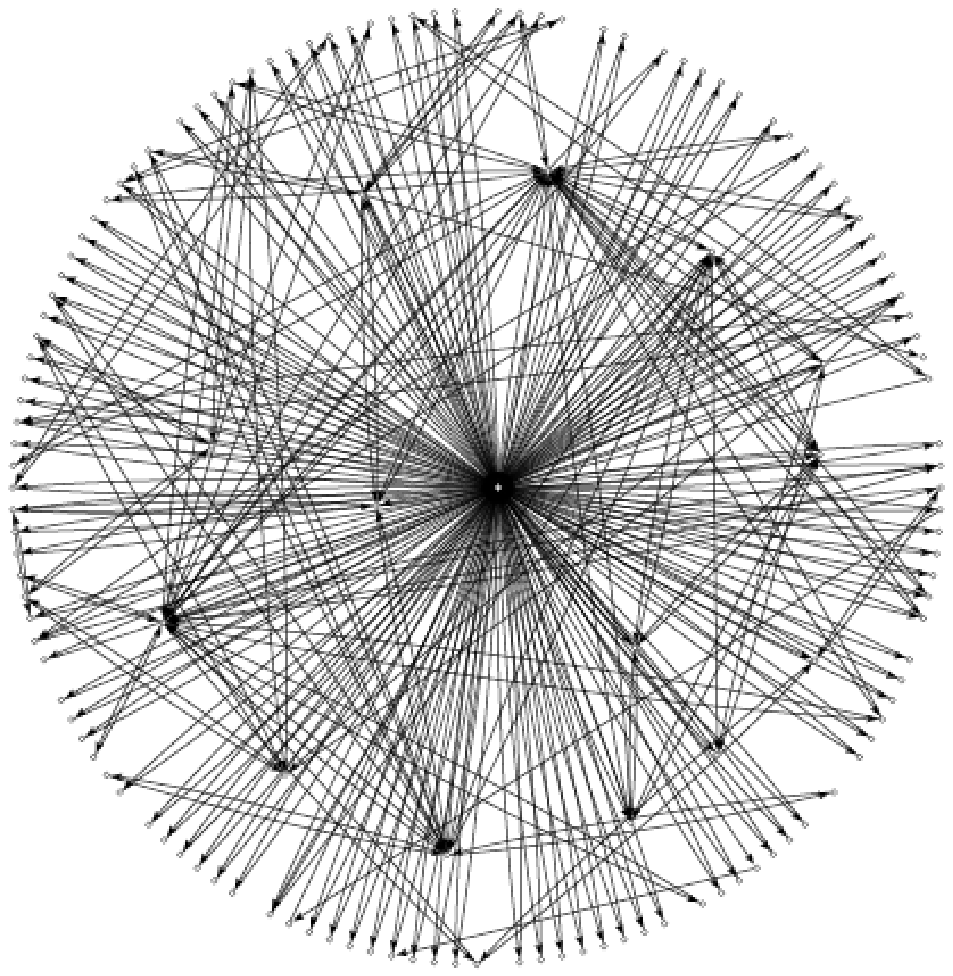}}
\resizebox{20.2pc}{!}{\includegraphics{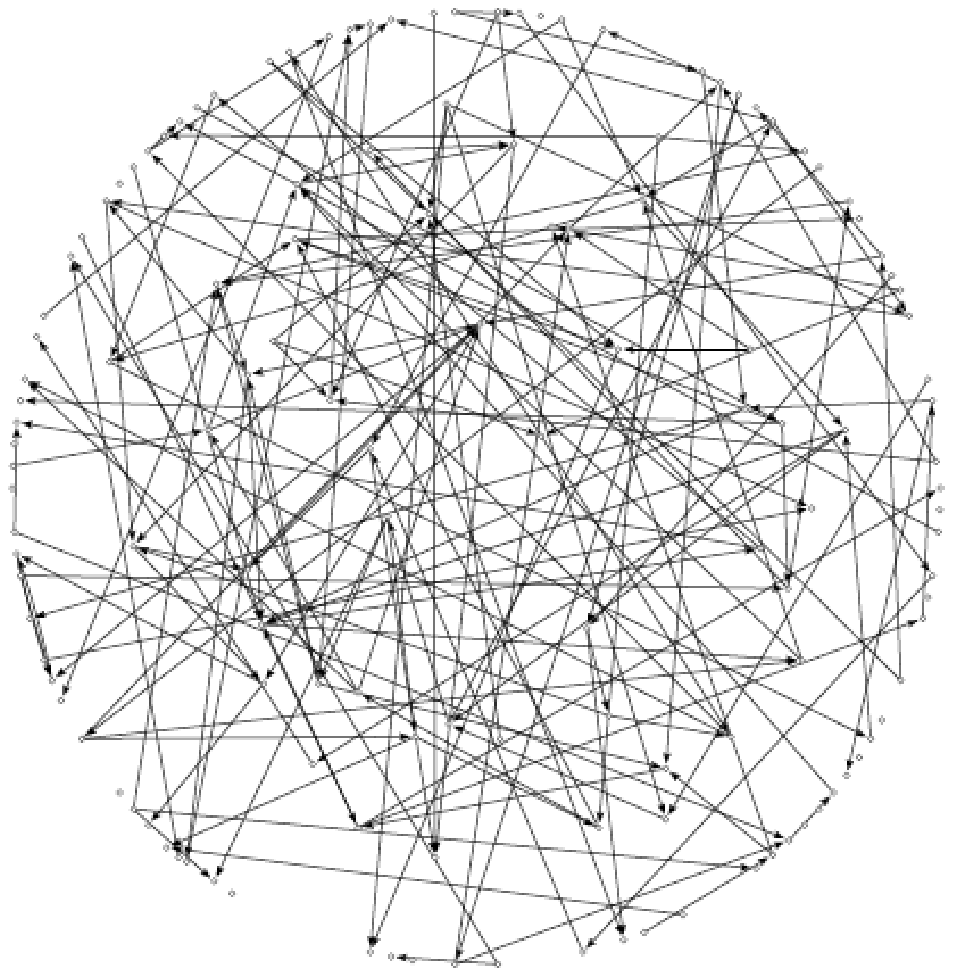}}
 \vspace{-.0truein}
\caption{\footnotesize Flight network structure of Austrian
Airline in a certain day (Friday).  Left panel: real network;
Right panel: Erd\"os and R\'enyi network with $N$ = 136 and
$\langle k \rangle$ = 1.31. Vertices represent the airports and
lines means connected flights. } \label{fig_netw}
\end{figure}

\begin{figure}
\resizebox{13.5pc}{!}{\includegraphics{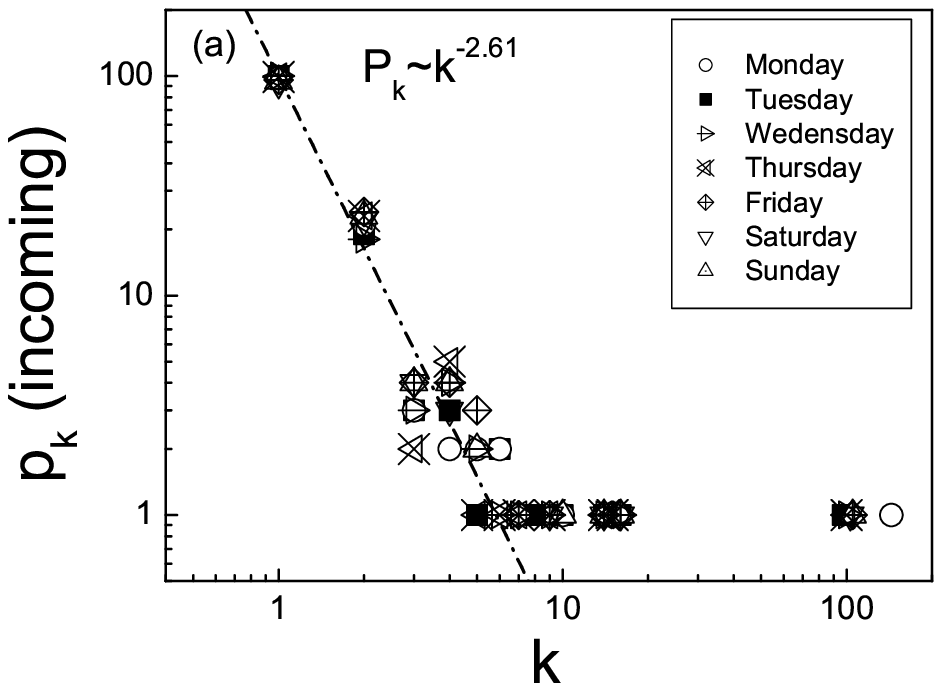}}
\resizebox{13.5pc}{!}{\includegraphics{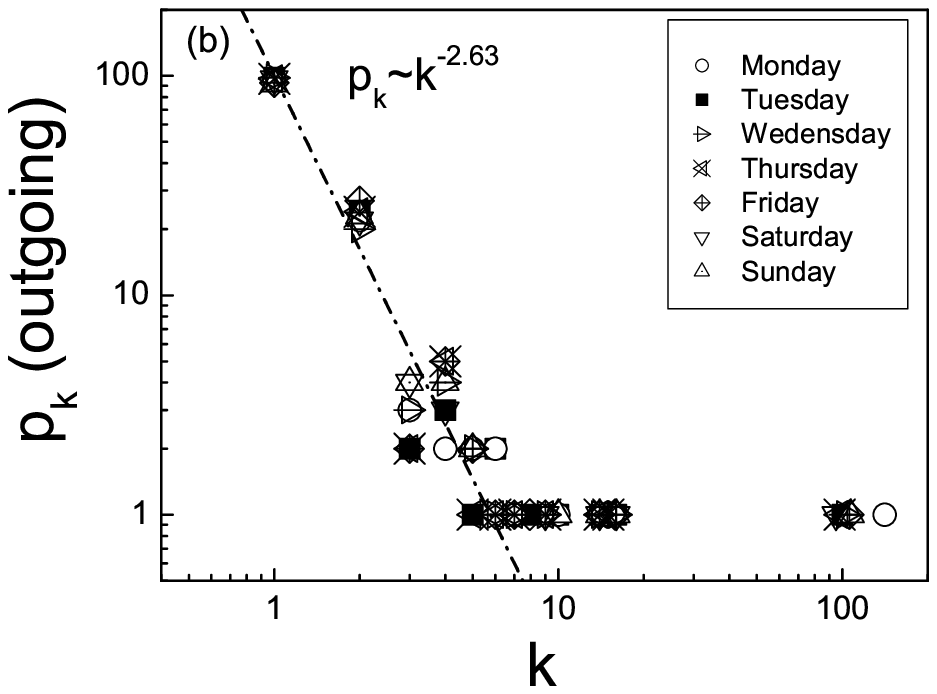}}
\resizebox{13.5pc}{!}{\includegraphics{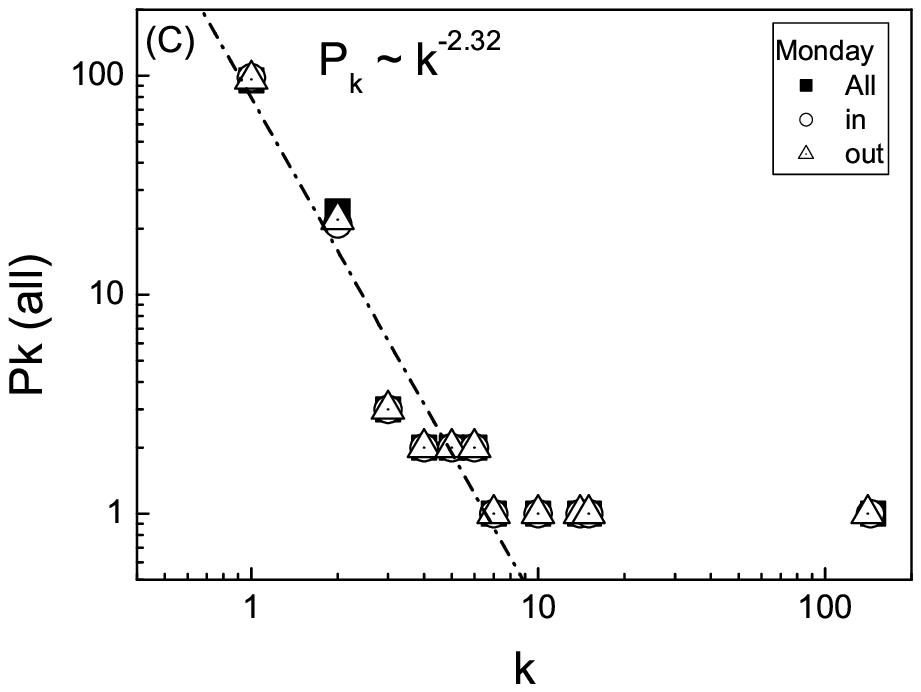}}
\vspace{-.2truein} \caption{\footnotesize Degree distribution for
each day during a week. (a): the case of incoming flights; (b):
the case of outgoing flights; (c) the case of all flights on
Monday. } \label{fig_degree}
\end{figure}
\end{widetext}

\begin{figure}
\includegraphics[scale=0.7]{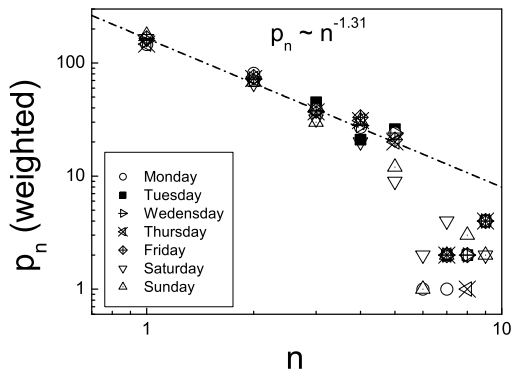}
\vspace{-0.3truein}\caption{\footnotesize Outgoing flight
distribution for weighted flight for each day during a week. }
\label{fig_weight}
\end{figure}

\begin{widetext}

\begin{figure}
\includegraphics[scale=1.5]{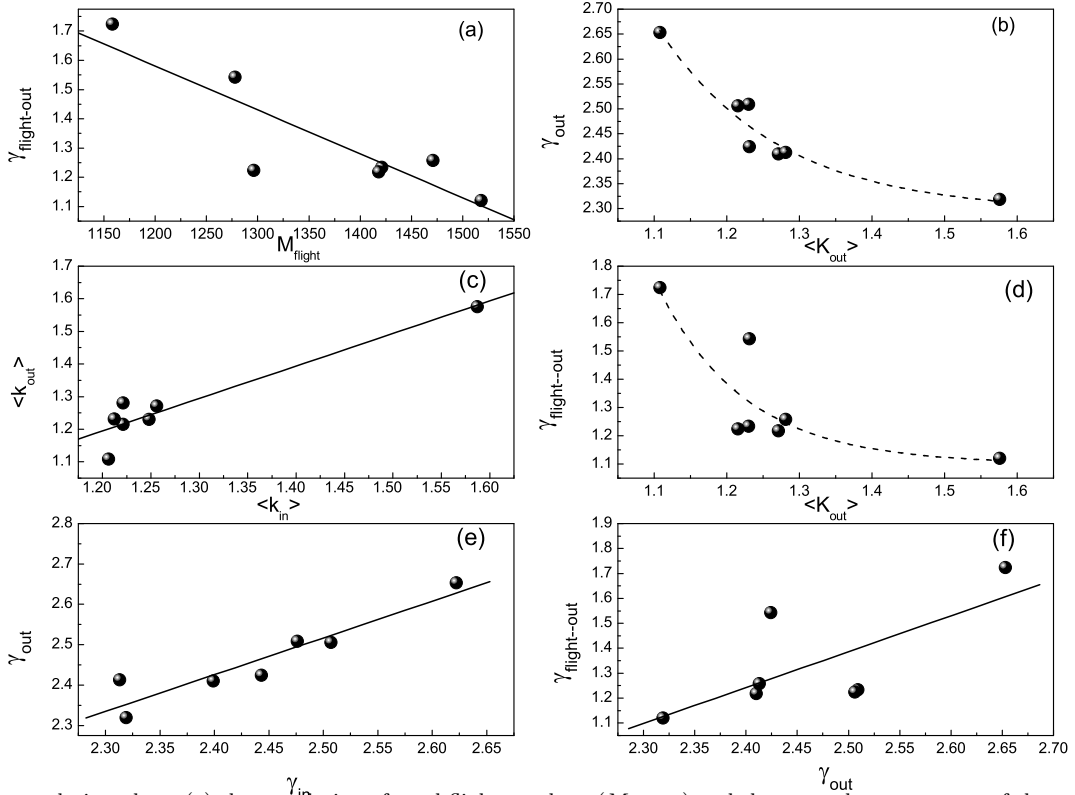}
\vspace{-0.6truein}\caption{\footnotesize  Some correlation plots:
(a) the correlation of total flight numbers ($M_{flight}$) and the
power-law exponents of the weight outgoing flight distributions
($\gamma_{flight-out}$); (b) the relationship between  $\langle
k_{out} \rangle $ and $\gamma_{out}$; (c) the correlation between
$\langle k_{in} \rangle$ and $\langle k_{out} \rangle$; (d) the
relationship between the $\langle k_{out} \rangle $ and
$\gamma_{flight-out}$; (e) the correlation between $\gamma_{in}$
and $\gamma_{out}$; (f) the correlation between $\gamma_{out}$ and
$\gamma_{flight-out}$. The solid lines represent the linear fits
and dashed lines for exponential decay fits. Each point represents
a day in a week (the } \label{fig_correl}
\end{figure}
\end{widetext}

\begin{table}[t]
  \centering
  \caption{ Comparison of relevant variables: (1) numbers of the airports $N$; (2)
  the numbers of fights $M$; (3)$\gamma$, which
represents the exponent of first segment of degree distribution;
(4) $ \langle k \rangle$, the average degree;
(5)$\gamma_{flight}$, the exponents of weight flight distribution;
(6) the clustering coefficient $C$ of the system; (7) the
assortative coefficient. } \label{TMF}
  \begin{tabular}{c c c c c c c c}
    \hline
      \multicolumn{1}{c}{} & \multicolumn{1}{c}{Mon} & \multicolumn{1}{c}{Tue} &
      \multicolumn{1}{c}{Wed} & \multicolumn{1}{c}{Thu} & \multicolumn{1}{c}{Fri} & \multicolumn{1}{c}{Sat} & \multicolumn{1}{c}{Sun} \\
  \hline
$N_{airport}$ & 133  & 136 &  134 &  136  &  136 & 130 &   134   \\
$M_{flight}$ & 1518  & 1421 &  1418 &  1296  &  1471 & 1158 &   1278   \\
$\gamma_{in}$ & 2.319  &  2.476 &  2.399 &  2.507  &   2.312 &   2.622 & 2.443  \\
$\gamma_{out}$ & 2.319  &  2.509 &  2.410 &  2.506  &   2.413 &   2.653 & 2.424  \\
$\gamma_{all}$ & 2.331  &  2.319 &  2.478  &  2.519  &   2.495  &   2.649 & 2.457  \\
$\langle k_{in} \rangle$ &  1.587  &   1.248 & 1.256  &  1.221  &  1.221  &  1.206 & 1.212  \\
$\langle k_{out} \rangle$ &  1.576  &   1.230 & 1.271  &  1.215  &  1.281  &  1.108 & 1.256  \\
$\langle k_{all} \rangle$ &  1.609  &   1.221 & 1.306  &  1.250  &  1.309  &  1.177 & 1.231  \\
$\gamma_{flight-out}$ &  1.120 &  1.234 & 1.218  &  1.224  &  1.258  &  1.724 & 1.543  \\
$C$ &  0.202  &   0.204 & 0.195  &  0.206 &  0.242  & 0.180 & 0.210  \\
$r$ & -0.529 & -0.515 & -0.519 & -0.517 & -0.517 & -0.562 & -0.543 \\
       \hline
  \end{tabular}
\end{table}

As shown in Figure 1, the topological structure of flight network
is significantly different from those of random graphs. In a
random graph of the type studied by Erd\"os and R\'enyi, each edge
is present or absent with equal probability, and hence the degree
distribution is binomial or Poisson distribution in the limit of
large graph size. Real-world networks are mostly found to be very
unlike the random graph in their degree distributions. The degrees
of the vertices in most network are highly right-skewed. This is
the case of the present flight network. From the exponents
$\gamma$ of different days in a week as shown in Table I, we can
find that exponents $\gamma_{in,out}$ and $\gamma_{all}$ on
Saturday are the largest and on Monday are basically the smallest.
Similarly, the mean degrees of flights on Saturday and Monday are
significantly different: it is the smallest on Saturday and the
largest on Monday, which is in consistent with the largest value
of $\gamma$ on Saturday and the smallest on Monday. This is also
not contradicted with the difference of total day-flight number
between Saturday and Monday as shown in the same table. In other
words, Monday is the busiest flight transportation day and
Saturday is the  most unoccupied flight transportation day for the
Austrian Airline. This can be partially related to the behavior of
human business travel.

Since the flight network involves in transportation flux, the
weight is important and can reflect some information of the whole
network. As shown in Fig.~\ref{fig_weight}, the  flight weight
distribution in a week has a power-law distribution in the small
weight branch,
\begin{equation}
P_n \sim n^{-\gamma_{flight}},
\end{equation}
where $n$ is the exact number of flights between any given airport
$i$ and $j$. The outgoing network exponents,
$\gamma_{flight-out}$,  of different days in a week are shown in
Table I. The mean exponent of a week is 1.33. Again,  there is a
different value for working days and weekend. The exponent
$\gamma_{flight-out}$ is around 1.2 in working days but it shifts
dramatically to larger values on Saturday and Sunday (see Table
I). Again, the value of Saturday is the largest and the one of
Monday is the smallest. The larger $\gamma_{flight-out}$ means the
steep slope, which results in the smaller mean weight. Therefore,
the values of larger $\gamma_{flight-out}$ can be attributed to
the declining flight number  on weekends.

To have  a visual feeling of the Table, we take the outgoing
flight network as an example to make some correlation plots which
are shown in Fig.~\ref{fig_correl}. Each point represent the data
of a day in one week.  Fig.~\ref{fig_correl}(a) shows the
correlation of total flight numbers ($M_{flight}$) and the
power-law exponents of the weight flight distributions
($\gamma_{flight}$). It is cleanly seen that the larger the total
flight numbers, the smaller the exponents, i.e. the flatter the
weight flight distribution. Fig.~\ref{fig_correl}(c) and (e)
display the positive correlation between $\langle k_{in} \rangle$
and $\langle k_{out} \rangle$, and $\gamma_{in}$ and
$\gamma_{out}$, respectively. This just says there is similar
behavior for both incoming flight network  and outgoing flight
network. Fig.~\ref{fig_correl}(f) shows the positive correlation
between $\gamma_{out}$ and $\gamma_{flight-out}$, i.e. there
exists the positive correlation between the degree distribution
and weight flight distribution. Fig.~\ref{fig_correl}(b) and (d)
shows the relationship between the $k_{out}$ and $\gamma_{out}$ or
$\gamma_{out}$, respectively, which indicates that the degree  or
weight flight distribution which has larger mean value of the
degree or weight flights has flatter distributions. From these
correlation plots, we found that the flight network shows its
evolving feature day by day in a week, i.e. there are different
exponent values of network distributions in different week-days
which are apparently related to total flight number of each
week-day.

\begin{figure}
\includegraphics[scale=0.7]{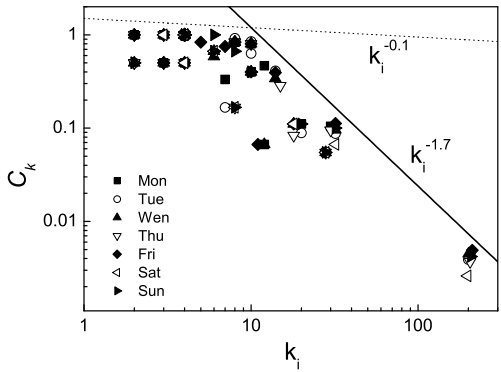}
\vspace{-0.3truein}\caption{\footnotesize The clustering
coefficient as a function of the vertex degree.  The dot line has
slope -0.1 and solid line has slope -1.7.} \label{fig_kici}
\end{figure}

In many social networks, there exists a clique form which can be
represented by circles of friends or acquaintances.
Quantitatively, this inherent tendency to cluster can be expressed
by the clustering coefficient \cite{Wat}. For a selected vertex
$i$ of the network, it has $k_i$ edges which we call the nearest
neighbours of $i$. In this case, the maximum possible edges among
$k_i$ neighbours are $k_i(k_i - 1)$ = 2. If we use $N_{real}$ to
denote the number of edges that actually exist, the clustering
coefficient of vertex $i$ can be written as
\begin{equation}
C_k = \frac{ N_{real}} { k_i (k_i - 1)/2}
\end{equation}
and the clustering coefficient of the entire network is defined as
$C = \frac{1}{ N} \sum_i C_k$.
The clustering coefficient $C$ of the Austrian Airline flight
 network in a week is 0.206. We also calculate $C$ on each day (see
Table 1). To look for the difference, we complete $C$ of our
 flight network with that of a random graph which has
the same $N$ and $\langle k \rangle$. In such a random graph, the
clustering coefficient is $C_{rand} = \langle k \rangle/N = p =
0.01$ where $p$ is the connection probability. Thus, $C$ in our
flight network is much larger than that in a random graph.
Fig.~\ref{fig_kici} shows the scattering plots of  the clustering
coefficient for un-directional flight network of each day in a
week as a function of the vertex degree. Similar to the degree
distribution (Fig.2), there are two segments: a nearly flat
distribution for small $k$ ($<7$) and a power-law decay with the
exponent $\sim 1.7$ for $k \geq 7$. The small-$k$ branch
corresponds to the majority of airports with a few links to other
airports, each such airport $i$ has a clustering coefficient close
to 1. The high-$k$ airports include many large airports, and thus,
their neighbors are not necessarily linked to each other,
resulting in a smaller $C_k$. A power-law decay of high-$k$ branch
indicates that a hierarchical organization \cite{Ravasz} for
larger airports, in contrast to the $k$ independent $C_k$
predicted by the scale-free networks as in small-$k$ branch.

\begin{figure}
\includegraphics[scale=0.7]{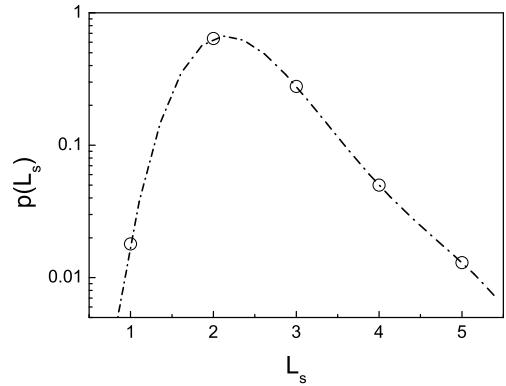}
\caption{\footnotesize Probability distribution of the  average
shortest-path length, $L_s$, of a certain day (Friday). }
\label{fig_D}
\end{figure}

The average shortest-path length between any two airports in the
system can be characterized by so-called "diameter"  in
small-world networks   \cite{BA2}, which is defined as
\begin{equation}
D = \frac{1}{ N(N -1)} \sum_{i \ne j} d_{ij},
\end{equation}
where $d_{ij}$ is the minimum number of edges traversed from
vertex $i$ to vertex $j$. The diameter of the flight network
reflects the average number of least possible connections between
any two airports. The corresponding probability distribution of
the shortest path lengths  of the whole flight network, namely 1,
2, 3, 4 and 5, is shown in Fig.~\ref{fig_D}, i.e. numerically
0.018, 0.641, 0.278, 0.050 and 0.013, respectively. The line is
just fourth order polynomial fit.
This implies that from airport $i$
to $j$, there will be basically not more than three connections
(the shortest-path length of 1 means a direct flight) where the
probability is smaller than 10$\%$. The diameter of our flight
network is $D$ = 2.383, which means that on the average there will
be 1.383 connections from airport $i$ to $j$. Using the same
approach, we compare the $D$ of our flight network with that of
the random graph. The diameter of the random graph is $D_{rand}$ =
ln(N)/ln($\langle k\rangle$) = 18.67 \cite{Koc}. In other word,
the diameter of our flight network is significantly smaller than
the one of the random graph with the same nodes and mean degree.

Many networks show ¡®assortative mixing¡¯ on their degrees, i.e.,
a preference for high- degree vertices to attach to other
high-degree vertices, while others show disassortative
mixing¡ªhigh-degree vertices attach to low-degree ones.
Quantitatively, the degree-degree correlation coefficient (also
called assortative coefficient)  can be written as
\begin{equation}
r = \frac{\frac{1}{M} \sum _i j_i k_i - [\frac{1}{M} \sum _i
\frac{1}{2}(j_i + k_i)]^2}{\frac{1}{M} \sum _i \frac{1}{2}(j_i^2 +
k_i^2) - [\frac{1}{M} \sum _i \frac{1}{2}(j_j+k_i)]^2},
\end{equation}
where $j_i$ and $k_i$ are the degrees of the vertices at the ends
of the $i$th edge, with $i$ =1,..., M.  As Newmann showed, the
values of $r$ of the social networks have significant assortative
mixing. By contrast, the technological and biological networks are
all disassortative \cite{Newmann}. In this work, we also check the
coefficient $r$ and we list those values of each day in Table 1.
As we expected the values are all negative, which means the flight
network is disassortative. In other word, the large airports are
likely to link to smaller airports. This fact is in agreement with
many technological and biological networks \cite{Newmann}. The
value of Saturday shows the largest.

In summary, our analysis demonstrates that  the Austrian Airline
flight network  displays the small-world property: high clustering
coefficient and small diameter. The clustering coefficient $C$
(0.206) is greatly larger than that of a random network (0.01)
with the same $N$ and $\langle k \rangle$ while the diameter $D$
(2.383) of the flight network is significantly smaller than the
value of the same random network (18.67). The degree distributions
for smaller airports show the power-law behavior with an exponent
2.47 for undirected flight networks. Also the flight weight
distributions have power-law distributions with exponent of
approximately 1.33. Further, the network  shows disassortative
behavior which indicates that  the large airports are likely to
link to smaller airports. In addition, the power-law decay
behavior of the clustering coefficient  against the degree for
high-$k$ nodes reflects that the large airports reveal the
hierarchical organization.  In a whole week, the power-law
exponents of the degree distribution and the flight weight
distribution show different values day-by-day, especially between
Monday and Saturday. The smallest exponent for Monday corresponds
to the busiest flight transportation day and the largest exponent
on Saturday corresponds to the most unoccupied flight
transportation day for the Austrian Airline.

The work was partially Supported by NSFC under Grant No. 10610285.
{}
\end{document}